\documentclass[12pt]{article}
\usepackage{graphics}
\usepackage{graphicx}

\input{epsf}
\usepackage{cite}
\def\@fmsl@sh#1#2#3{\m@th\ooalign{$\hfil#1\mkern#2/\hfil$\crcr$#1#3$}}
 \def\eq#1\en{\begin{equation}#1\end{equation}}
\def\s[#1,#2]{[#1\stackrel{\star}{,}#2]}
\def\sx[#1,#2]{[#1\stackrel{\star_{x}}{,}#2]}

\newcommand{\nc}{\newcommand}
\nc{\beq}{\begin{equation}}
\nc{\eeq}{\end{equation}}
\nc{\beqa}{\begin{eqnarray}}
\nc{\eeqa}{\end{eqnarray}}

\def\gsim{\mathrel{\rlap{\lower4pt\hbox{\hskip1pt$\sim$}}
    \raise1pt\hbox{$>$}}}       

\begin{document}
\makeatletter
\def\fmslash{\@ifnextchar[{\fmsl@sh}{\fmsl@sh[0mu]}}
\def\fmsl@sh[#1]#2{%
  \mathchoice
    {\@fmsl@sh\displaystyle{#1}{#2}}%
    {\@fmsl@sh\textstyle{#1}{#2}}%
    {\@fmsl@sh\scriptstyle{#1}{#2}}%
    {\@fmsl@sh\scriptscriptstyle{#1}{#2}}}
\def\@fmsl@sh#1#2#3{\m@th\ooalign{$\hfil#1\mkern#2/\hfil$\crcr$#1#3$}}
\makeatother


\title{\large{\bf Bound on  four-dimensional Planck mass}}

\author{Xavier~Calmet\thanks{Charg\'e de recherches du F.R.S.-FNRS} \thanks{xavier.calmet@uclouvain.be } and Marco~Feliciangeli\thanks{marco.feliciangeli@uclouvain.be}  \\
Catholic University of Louvain \\
Center for Particle Physics and Phenomenology\\
2, Chemin du Cyclotron\\
B-1348 Louvain-la-Neuve, Belgium
}

\date{June, 2008}

\maketitle

\begin{abstract}
  In this note we derive a bound using data from cosmic rays physics on  a model recently proposed to solve the hierarchy problem by lowering the Planck scale to the TeV region without the introduction of extra-dimensions. We show that the non observation of  small  black holes by AGASA  implies a model independent limit for the four-dimensional reduced Planck mass of roughly 488 GeV.
    \end{abstract}


\newpage

Different four dimensional models designed to address the hierarchy problem  which predict  strong scattering cross-sections in the tera-scale have recently been proposed \cite{Calmet:2008tn,Dvali:2001gx,Dvali:2007hz}. The most remarkable feature  of these models is the possible formation of quantum black holes in high energetic collisions of particles. The LHC which will start to operate in the coming months will be able to probe this energy domain, but cosmic rays experiments are already sensitive to strong interactions of cosmic rays neutrinos with nuclei in the Earth atmosphere. Anchordoqui {\it et al.} \cite{Anchordoqui:2001cg,Anchordoqui:2003jr} have derived bounds on the scale of quantum gravity for extra-dimensional models using data from AGASA. The aim of this work is to derive a similar limit on the scale of quantum gravity in four-dimensions following the work of Anchordoqui {\it et al.} very closely. We shall first summarize the model proposed in \cite{Calmet:2008tn}.

The strength of the gravitational interaction is renormalized by matter field fluctuations \cite{Larsen:1995ax,Veneziano:2001ah,Calmet:2008tn}. One finds that the effective Planck mass depends on the energy scale $\mu$  as
\begin{eqnarray}
\label{Mrunning}
M( \mu )^2 ~=~
M(0)^2 - \frac{\mu^2}{12 \pi} \left(N_0+N_{1/2}-4 N_1\right),
\end{eqnarray}
where $N_0$, $N_{1/2}$ and $N_1$ are the numbers of real spin zero scalars, Weyl spinors and spin one gauge bosons coupled to gravity and where $M(0)=1.22090 \times 10^{19}$ GeV. Related calculations have been performed in string theory and lead to the same behavior for the running of the Planck mass \cite{Kiritsis:1994ta}.

If the strength of gravitational interactions is scale dependent, the true scale $\mu_*$ at which quantum gravity effects are large is the one at which
\begin{equation}
\label{strong}
M (\mu_*) \sim \mu_*.
\end{equation} 
This condition means that fluctuations in spacetime geometry at length scales $\mu_*^{-1}$ will be unsuppressed. It has been shown in \cite{Calmet:2008tn} that the presence of a large number of fields can dramatically impact the value $\mu_*$. For example, it takes $10^{32}$ scalar fields and or Weyl spinors to render $\mu_* \sim {\rm TeV}$, thereby removing the hierarchy between weak and gravitational scales. The most striking feature of this model is that small black holes will form in particles collisions with center of mass energies of the order of 1 TeV. We shall now derive a bound on the scale of quantum gravity that apply to this model but also more generically to quantum gravity in four-dimension and thus use $M_P$ instead of $\mu_*$ in the sequel. 

Following the work of Anchordoqui {\it et al.} \cite{Anchordoqui:2001cg,Anchordoqui:2003jr} , we use the observation of quasi-horizontal showers by AGASA  \cite{Yoshida:2001pw,Inoue:1999cn} which translates into an upper bound on the number of small black holes of 3.5 \cite{Anchordoqui:2001cg} produced during the run time $T=1710.5$ days of the experiment. These small black holes would be produced in collisions of high energetic Earth-skimming neutrinos with nuclei in the Earth atmosphere. The cross-section $\nu$ N $\to$ BH is given by
\begin{equation}
\label{CSection}
\sigma(E_\nu,x_{min},M_{R})=
\int^1_0 2z dz \int^1_{\frac{( x_{min} M_{R})^2}{y(z)^2 s_{max}}} dx F(4) \pi r_s^2(\sqrt{\hat s},M_{R}) \sum_i f_i(x,Q)
\end{equation} 
where $M_R=M_P/\sqrt{8\pi}$ is the reduced Planck mass, $x_{min}=M_{BH}^{min}/M_R$ is the ratio of the minimal black hole mass which can be created to the reduced Planck mass, $F(4)$ is the Eardley  Giddings correction which describes the fact that not all of the energy of the partons is available for black hole formation \cite{Eardley:2002re}, $y(z)$ is the inelasticity function calculated in \cite{Yoshino:2002tx} following the work of Eardley and Giddings  \cite{Eardley:2002re}, $\hat{s}= 2 x m_N E_{\nu}$ where $m_N$ is the  nuclei mass and $E_{\nu}$ is the neutrino energy. The functions $f_i(x,Q)$ are the parton distribution functions (we use CTEQ5 for which an unofficial mathematica version is available on the webpage of the CTEQ collaboration). We take $Q\sim 1000$ GeV, as noted in \cite{Anchordoqui:2001cg}, the choice of $Q$ does not impact much the outcome of the calculation. Finally $r_s$ is the Schwarzschild radius and is given by
\begin{equation}
\label{SR}
 r_s(\sqrt{\hat s},M_{R})= \frac{ \sqrt{\hat s}}{4 \pi M_R^2}.
\end{equation} 
 The black holes produced in the reaction $\nu$ N $\to$ BH can be charged under U(1), SU(3) but they could in principle also be neutral under these two gauge symmetries. We shall consider both semi-classical black holes (for which the construction of Eardley and Giddings applies i.e. $x_{min} \ge 3$) and what we call quantum black holes \cite{Calmet:2008tn} ($x_{min} \ge 1$ which only decay to a couple of particles. The three particles final state is strongly suppressed with respect to the two particles final state because of phase space.  Because gravity conserves gauge charges, most quantum black holes, created in collisions of particles charged under SU(3)$\times$U(1)$_{em}$, will principally decay to standard model particles and not to the large hidden sector ($10^{32}$ particles) which would make them invisible. This is important for collider experiments. However, in the case of AGASA which is sensitive to a suppression of the neutrino flux due to new strong interactions between neutrinos and nuclei, it is not important if black holes decay visibly or invisibly and we can thus sum over all the possible intermediate black holes.
 
 Following  \cite{Anchordoqui:2001cg}  we consider the flux of guaranteed cosmogenic GZK neutrinos which originate from the collision of high energy protons on the cosmic microwave background photons producing a delta resonance which then decays to a charged pion among other particles. This pion  decays to a lepton and a neutrino. This is the famous GZK mechanism for the suppression of the spectrum of high energetic cosmic rays above $10^{19}$ eV.
 The number of black holes expected is given by
 \begin{equation}
\label{number}
N(M_{R})= N_A T \int dE_\nu \int^1_0 2z dz \int^1_{\frac{( x_{min} M_{R})^2}{y(z)^2 s_{max}}} dx  \frac{d\Phi}{dE_\nu} A(y E_\nu)F(4) \pi r_s^2(\sqrt{\hat s},M_{R}) \sum_i f_i(x,Q)
\end{equation} 
where  $\frac{d\Phi}{dE_\nu}$ is the flux of cosmic neutrinos and $A(y E_\nu)$ is the acceptance of the experiment under consideration. We fit both functions to the plots given in \cite{Anchordoqui:2001cg} in figures 2 and 4. We choose to use the neutrino flux corresponding to the work of Protheroe and Johnson \cite{Protheroe:1995ft}. Note that there is some model dependence both in the neutrino flux and in the acceptance and the bounds derived from AGASA are thus more order of magnitude estimates rather than tight bounds.  Furthermore, as discussed above, we take $x_{min}=1$.

It is worth mentioning that the mechanism proposed by Lykken {\it et al.} \cite{Lykken:2007kp}  who  have studied the possibility that neutrinos would annihilate through gravitational interactions with e.g. supernovae neutrinos on their way to Earth does not yield a sizable suppression of the GZK neutrino flux in our case. The gravitational interaction is too weak to suppress the flux in a sizable manner. Similarly the GZK production mechanism for the neutrinos is not affected by the new gravitational interaction.

Requesting that $N(M_{R}) < 3.5$ black holes, we find a bound on the scale of four-dimensional quantum gravity $M_R> 488$ GeV (i.e. for the reduced Planck mass) and $M_P> 2.4$ TeV for the Planck mass. It is remarkable that this bound is independent on the details of the model proposed in \cite{Calmet:2008tn}. It is also independent on assumptions about quantum gravity such as possible violation of symmetries, e.g. violation of Lorentz invariance, which leads to much tighter bounds (see e.g. \cite{Jacobson:2004rj}). We note that our bound also applies to the model proposed in \cite{Calmet:2007je} where gravity remains weak, but where a new scalar similar to a dilaton which stabilizes the Planck mass can lead to strong rescattering effects. It is worth mentioning that if the inelasticity function and the Eardley Giddings factor are set to unity, one finds a tighter bound of 565 GeV for the reduced four-dimensional Planck scale. Finally, we point out that as mentioned in  \cite{Anchordoqui:2001cg}  the acceptance of the Pierre Auger Observatory is typically bigger than that of AGASA by a factor 30 and this experiment should thus be able to push the bounds on the Planck scale by a factor 2 (for a comparable run time). The limit on the  scale of four dimensional quantum gravity is so  weak that quantum black holes could be just around the corner, i.e. if they play a role in the resolution of the hierarchy problem, and are certainly within the reach of the LHC. A study of quantum black holes at the LHC is in preparation and will appear shortly \cite{CWH}.

\subsection*{Acknowledgments}

We would like to thank Jean-Marc Gerard, Eduardo Cortina Gil, Stephen Hsu, Cecilia Lunardini and Fabio Maltoni for helpful discussions. This work is supported in part by the Belgian Federal Office for Scientific, Technical and Cultural Affairs through the Interuniversity Attraction Pole P6/11.

\noindent


\bigskip


\end{document}